\documentclass[11pt,onecolumn,draftclsnofoot]{IEEEtran}
\usepackage{psfrag,amsmath,epsfig,amssymb,array,setspace, graphics, graphicx, url, amsfonts,color}
\usepackage{booktabs}
\usepackage{subfigure}
\usepackage{cite}
\def \Re    {\operatorname{Re}}
\def \Im    {\operatorname{Im}}
\def \Diag      {\operatorname{Diag}}
\def \tr    {\textrm{tr}}

\def \SINR  {\textrm{SINR}}
\def \vec   {\textrm{vec}}

\def \Diag  {\textrm{Diag}}
\def \mse   {\textrm{MSE}}

\def \I     {{\mathbf I}}

\def \P     {{\mathbf P}}
\def \A     {{\mathbf A}}

\def \B     {{\mathbf B}}
\def \C     {{\mathbf C}}

\def \Q     {{\mathbf Q}}

\def \Pa    {\underline{\mathbf P}}
\def \Ba    {\underline{\mathbf B}}

\def \Ptot  {P_{\textrm{total}}}

\def \signk     {\sigma_{n_{k}}^{2}}

\def \sigv  {\sigma_{v}^{2}}

\def \0     {{\mathbf {0}}}
\def \e     {{\mathbf e}}
\def \s     {{\mathbf s}}

\def \e     {{\mathbf e}}
\def \x     {{\mathbf x}}

\def \b     {{\mathbf b}}
\def \u     {{\mathbf u}}

\def \e     {{\mathbf e}}
\def \v     {{\mathbf v}}
\def \u     {{\mathbf u}}
\def \i     {{\mathbf i}}
\def \a     {{\mathbf a}}
\def \b     {{\mathbf b}}

\def \m     {{\mathbf m}}
\def \h     {{\mathbf h}}
\def \w     {{\mathbf w}}

\def \ha    {\underline{\mathbf h}}

\def \ea    {\underline{\mathbf e}}
\def \ba    {\underline{\mathbf b}}
\def \ma    {\underline{\mathbf m}}

\def \phib  {{\boldsymbol {\phi}}}

\def \Hb    {\hat{\mathbf{H}}}

\def \sh    {\hat{\mathbf{s}}}
\def \hh    {\hat{\mathbf{h}}}
\def \hha   {\hat{\underline{\mathbf{h}}}}

\def \Q     {\mathbf {Q}}

\newtheorem{lem}{Lemma}
\newtheorem{Design_Formulation}{Design Formulation}

\newcommand{\revision}[1]{{\color{black} #1}}
\begin{document}
%------------------------------------------------------------------------------------------------------------------
\title{Non-linear and Linear Broadcasting  with MSE Constraints: Tractable Approaches for Scenarios with Uncertain Channel Information }
%------------------------------------------------------------------------------------------------------------------
\author{Michael Botros Shenouda,~\IEEEmembership{Student Member,~IEEE,} and
        Timothy N. Davidson,~\IEEEmembership{Member,~IEEE,}
%----------------------------------------------------------------------------------------------------------
\thanks{This work was supported in part by an Ontario Graduate Scholarship, and by the Natural Sciences and Engineering Research Council of Canada.
The work of the second author is also supported by
the Canada Research Chairs Program.
} % End of first \thanks
%
%----------------------------------------------------------------------------------------------------------
%\thanks{
%A preliminary version  of portions of this manuscript  appear\revision{s} in the \textit{Proceedings of the $20^{\text{th}}$ Canadian Conference on Electrical and Computer Engineering}, Vancouver, April 2007.
%}% End of second \thanks
}% End of \author
\date{}
%----------------------------------------------------------------------------------------------------------
% The paper headers
\markboth{
Draft
}
{Shell \MakeLowercase{\textit{Botros Shenouda et al.}}:
Non-linear and Linear Broadcasting  with MSE Constraints}
\maketitle
%------------------------------------------------------------------------------------------------------------------ 
%----------------------------------------------------------------------------------------------------------------------
\begin{abstract}
%-------------------------------------------------------------------------------------------------------------------
We consider the downlink of cellular systems in which the base station employs multiple transmit antennas and 
each user has one receive  antenna.
We consider communication schemes in which the users have Quality of Service (QoS) requirements, and we study the design of 
robust broadcasting schemes that minimize the transmission power necessary to guarantee the QoS requirements for all channels within bounded uncertainty regions around the transmitter's estimate of  each user's channel.
We formulate each user's QoS requirement as a constraint on the  mean square error (MSE) in its received signal, and we show that these MSE  constraints imply constraints on the received  signal-to-interference-plus-noise-ratio (SINR) of each user.
Using these MSE constraints, we present a unified design approach for robust linear and non-linear transceivers with QoS requirements. 
The proposed designs overcome the limitations of existing approaches that only provide  conservative solutions and only applicable to the case of linear precoding.
Furthermore, we provide tractable and computationally-efficient design formulations for a quite general model of channel uncertainty that subsumes many uncertainty regions.
we also consider the problem of robust counterparts to precoding schemes that maximizes the weakest user's signal subject to a total power constraint on the transmitting antennas.
For this problem, we provide  quasi-convex formulations, for both non-linear and linear transceivers, that can be efficiently solved using a one-dimensional bisection search. 
Our numerical results demonstrate that in the presence of uncertainty in the transmitter's knowledge of users' channels, the proposed designs provide  guarantees to a larger set of QoS requirements than existing approaches, and require less transmission power to satisfy these requirements.
%-------------------------------------------------------------------------------------------------------------------
\end{abstract}

\newpage
\section{Introduction}
%---------------------------------------------------------------------------------------------------------------------------------------------
% By the prayers of St. Kirolos and St. Mina, and first and last the prayers of St. Mary.
%---------------------------------------------------------------------------------------------------------------------------------------------
%----------------------
% Perfect CSI
%----------------------
The design of wireless broadcasting schemes that satisfy the quality of service (QoS) requirements of the intended users (receivers)
is of growing interest in interactive communication applications  and in the downlink of cellular systems with differentiated services.
The provision of multiple antennas at the transmitter (base station) of the downlink enables the design of schemes that (attempt to) satisfy the users' QoS requirements by spatially precoding the users' data in order to mitigate the multiuser interference at the (disjoint) receivers.
The availability of accurate channel state information (CSI) at the transmitter is important in such schemes as it enables the mitigation of the interference experienced by the receivers as a result of channel propagation.
For scenarios in which one can assume perfect CSI is available at the transmitter, the problem of designing a  precoder  that minimizes the transmitted power required to satisfy a set of QoS requirements has been considered in  \cite{Rashid-Farrokhi_1998_COM, Rashid-Farrokhi_1998_JSAC, Bengtsson_1999_Allerton,Beamforming_Bengtsson_2001,Sol_SINR_Schubert_2004,Wiesel_2006_Fixed} for the case of linear precoding, and in \cite{THP_Independent_MSE,Doostnejad_2005_QoS,THP_Schubert_MSE,Sanguinetti_2007_QoS,Schubert_2005_QoS_NL} for the case of non-linear precoding.

%----------------------
% Imperfect CSI
%----------------------
In practical broadcasting systems, the CSI that is available at the transmitter is subject to a variety of  sources of imperfection,  such as estimation errors, channel quantization errors and short channel coherence time.
For example, in communication scenarios in which the receivers  feed back their quantized CSI to the transmitter (e.g., \cite{Jindal_2005_BC_LimitedFB,Ding_2007_BC_LF,Yoo_2007_LF}), the uncertainty in the CSI that is available at the transmitter is dominated by quantization errors.
Downlink precoder design methods that assume perfect CSI are particularly sensitive to these uncertainties, which can result in serious degradation of the quality of the received signals  \cite{Jindal_2005_BC_LimitedFB,Ding_2007_BC_LF}.
This suggests that the design of downlink precoding schemes should incorporate robustness to channel uncertainty.
One approach to incorporating   robustness  is  to consider  a bounded model for the error in the transmitter's estimate of the channels and to constrain the design the precoder so that the users' QoS requirements are satisfied for all channels admitted by this model.
This bounded uncertainty model is useful for systems in which it is difficult to provide the transmitter with an accurate statistical model for the channel uncertainty. In particular, it is useful for systems in which users feed back quantized channel measurements to the transmitter, as
knowledge of the quantization codebooks can be used to bound the quantization error.

For the downlink of cellular systems in which each receiver has a single antenna, the design of a linear precoder that minimizes the transmitted power required  to guarantee  that each user's QoS requirement is satisfied for all admissible channels was considered in \cite{Botros_Davidson_CCECE07b, Botros_QoS_JSTSP}; 
see also \cite{Bengtsson_1999_Allerton,Biguesh_2004_PL} for designs based on a bounded model for the errors in the transmitter's estimate of the  (deterministic) autocorrelation matrices of the channel.
These different approaches approaches formulated the QoS requirements as  constraints on the signal-to-interference-plus-noise (SINR) of each user.
While the methods proposed in \cite{Botros_Davidson_CCECE07b, Botros_QoS_JSTSP} provide tractable design formulations and significant improvements in performance over previous existing designs, those approaches have two limitations.
First, they are  not directly applicable to non-linear precoding schemes such as Tomlinson-Harashima  precoding (THP).
Second, when QoS is quantified in an SINR sense, the  robust linear QoS problem resulted in designs whose tractability is an open problem; see also \cite{Bertsimas_2006_App-RCO}.
In order to obtain tractable designs, a conservative design approach was taken in \cite{Botros_Davidson_CCECE07b, Botros_QoS_JSTSP}, and that approach requires the SINR constraints to be satisfied for a superset of the original bounded set of admissible channels.
%----------------------
% Our Work
%----------------------
In this paper, we provide remedies to both these limitations by providing tractable formulations (in the form of semidefinite programs) of both linear and non-linear downlink precoding schemes that minimize  the transmitted power required to ensure that each user's QoS requirement is satisfied for all admissible channels,  without expanding the admissible set.
We consider each user's QoS requirement as a constraint on the  mean square error (MSE) in each user's received signal, and we show these MSE  constraints imply constraints on the received  SINR of each user.
%The proposed approach provide more efficient designs by satisfying  the QoS constraints for the specified uncertainty set rather than an expanded set.
Since the QoS is measured in terms of MSE, our approach is applicable to non-linear Tomlinson-Harashima precoding and to linear precoding as a special case. In addition, the proposed designs (for the linear case) are obtained with lower computational cost cost than those based on SINR formulations of the QoS in \cite{Botros_Davidson_CCECE07b, Botros_QoS_JSTSP}.
Furthermore, we present a unified treatment of a quite general bounded uncertainty model that can represent uncertainty regions resulting from many quantization schemes.
The model naturally  includes channel uncertainty regions that are described using intersection of multiple uncertainty sets, e.g., interval constraints on the entries of each user's channel. 
While we provide exact robust design formulations for these types of uncertainties, we also provide conservative formulations that reduce the computational complexity of the design for these cases.

%----------------------
% Other two problems
%----------------------
The proposed design approaches can be extended to obtain efficiently-solvable quasi-convex formulations of some related design problems.
In particular, we consider  the robust counterpart of the problem of maximizing the weakest user's signal (minimizing the largest MSE among the users).
For precoding schemes that assume perfect CSI at the transmitter, this problem was studied for the case of linear precoding schemes in \cite{Sol_SINR_Schubert_2004,Wiesel_2006_Fixed}.
% and in \cite{} for non-linear TH precoding schemes.
For the bounded channel uncertainty model, tractable conservative approaches to the robust counterpart  of the linear  minimax precoder design problem were provided in \cite{Botros_QoS_JSTSP}, but the problem has remained open for the case of non-linear precoding.
We provide quasi-convex formulations of this robust minimax problem, for both non-linear and linear precoding schemes. These formulations can be efficiently solved using a one-dimensional bisection search.
%, in which a convex semidefinite program (SDP) is solved at each step.
We also show that this problem can be formulated as generalized   eigenvalue  problem; e.g., \cite{Boyd_1993_GEVP}.

We also consider the problem of determining the largest uncertainty region for which the QoS requirements can be satisfied for all admissible
channels using finite transmission power.
This problem is of considerable interest in the design of quantization codebooks for quantized channel feedback schemes.
In that case, one might wish to choose the rate of the channel quantization scheme to be large enough  (and the quantization cells small enough) for it to be possible to design a robust precoder with finite power.  We provide quasi-convex formulations of  this problem, too.
Our numerical results demonstrate the efficiency of the proposed approaches.
In particular, they provide  guarantees to a larger set of QoS requirements than existing approaches, and require less transmission power in order to satisfy these requirements.
%

%----------------------
% Our Notation
%----------------------
Our notation is as follows: We will use boldface capital letters to
denote matrices,  boldface lower case letters to denote vectors and
medium weight lower case letters to denote individual elements;
$\A^T$ and $\A^H$ denote the transpose and the conjugate transpose
of the matrix $\A$, respectively. The notation $\| \x \|$ denotes
the Euclidean norm of vector $\x$, while $\| \A \|$ denotes the
spectral norm (maximum singular value) of the matrix $\A$, and
$\mathrm{E} \{ \cdot \}$ denotes the expectation operator. The term
$\tr(\A)$ denotes the trace of matrix $\A$, $\A \otimes \B$ denotes
the Kronecker product of $\A$ and $\B$, and for symmetric matrices
$\A$ and $\B$, $\A \ge \B$ denotes the fact that $\A - \B$ is
positive semidefinite.
%---------------------------------------------------------------------------------------------------------------------------------------------

%--------------------------------------------------------------------------------------------------------------------
\section{System Model}
\label{QOS_MSE_Sec_Model}
%--------------------------------------------------------------------------------------------------------------------

\begin{figure}
\centering
\includegraphics[width=0.9\textwidth]{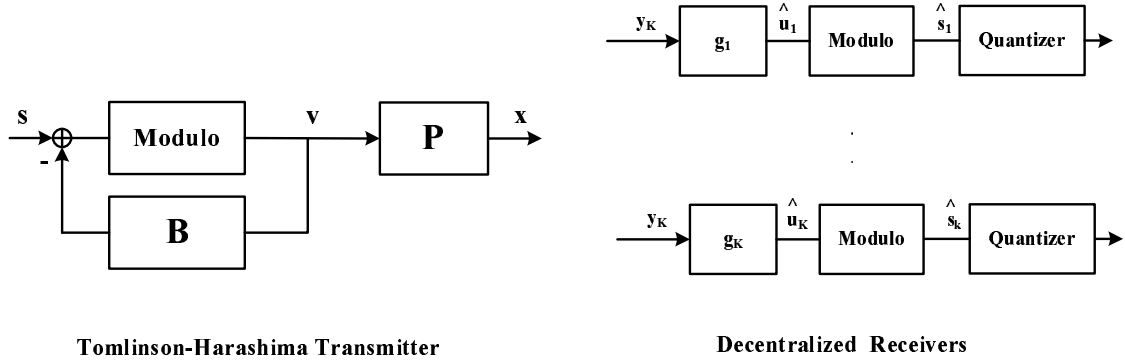}
\caption{Multiple-input single-output downlink system with Tomlinson-Harashima precoding at the transmitter.}
\label{QoS_MSE_Fig_Model_THP}
\end{figure}

%%%%%%%%%%%%%%%%%%%%%%%%%%%%%%%%
% THP precoding
%%%%%%%%%%%%%%%%%%%%%%%%%%%%%%%%
We consider the downlink of a multiuser cellular  communication system with $N_t$ antennas at the transmitter and $K$ users, each with one receive antenna.
We consider downlink systems in which Tomlinson-Harashima precoding (THP)  is used at the transmitter for multi-user interference pre-subtraction.
As shown in Fig. \ref{QoS_MSE_Fig_Model_THP},  the interference pre-subtraction and channel spatial equalization  at the transmitter can be modelled 
using a feedback  matrix $\B \in \mathbb{C}^ {K \times K}$ and a feedforward precoding matrix $\P \in \mathbb{C}^ {N_t \times K}$.
Since linear precoding is the special case of the THP  model in which $\B = \0 $, we will focus our development on the THP case and will extract the special case results for linear precoding as they are needed.
The vector $\s \in \mathbb{C}^{K}$ in Fig. \ref{QoS_MSE_Fig_Model_THP} contains the data symbol destined for each user, and we assume that $s_k$ is chosen from a square QAM constellation $\mathcal{S}$ with cardinality $M$  and that $\mathrm{E}\{\s \s^H\} = \I$.
The Voronoi region of the constellation $\mathcal{V}$ is a square whose side length is $D$.

\begin{figure}
\centering
\includegraphics[width=0.55\textwidth]{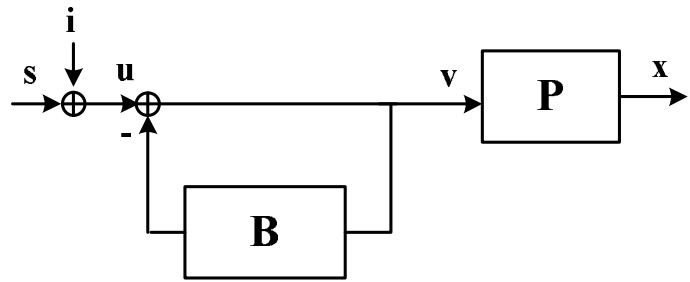}
\caption{Equivalent linear model for the transmitter.}
\label{QoS_MSE_Fig_Model_THP_2}
\end{figure}

In absence of the modulo operation,  the output symbols of the feedback loop in Fig.~\ref{QoS_MSE_Fig_Model_THP}, $v_k$, would be generated
successively according to  
$
v_k = s_k- \sum_{j = 1}^{k-1} \: \B_{k,j} v_j,   \label{QoS_MSE_THP_without_modulo}
$
where at the $k^{\text{th}}$ step, only the previously precoded symbols $v_1, .., v_{k-1}$ are subtracted.
Hence, $\B$ is a strictly lower triangular matrix.
%The summation in equation (\ref{QoS_MSE_THP_without_modulo})  suggests that the magnitude of $v_k$ may grow beyond the boundaries of $\mathcal{V}$.
The role of the transmitter's modulo operation is to ensure that $v_k$ remains within the boundaries of $\mathcal{V}$, and its effect is equivalent to the addition of the complex quantity $i_k = i_k^{{re}} \; D + j \; i_k^{{imag}} \; D$
to $v_k$, where $ i_k^{{re}}, \; i_k^{{imag}} \in \mathbb{Z}$, and $j = \sqrt{-1}$ .
Using this observation, we obtain the standard linearized model of the transmitter that does not involve a modulo operation, as shown in Fig. \ref{QoS_MSE_Fig_Model_THP_2}; e.g., \cite{F_THP_Book}. 
For that model,  
\begin{equation}
    \v = (\I + \B)^{-1} \u,       \label{QoS_MSE_Rel_v_u}
\end{equation}
where $\u = \i + \s$ is the modified data symbol.
As a result  of the modulo operation,  the elements of $\v$ are almost uncorrelated and uniformly distributed over the Voronoi region $\mathcal{V}$ \cite[Th. 3.1]{F_THP_Book}.
Therefore, the symbols of $\v$ will have slightly higher average energy than the input symbols $\s$. (This slight increase in the average energy is termed precoding loss \cite{F_THP_Book}.)
For example, for  square $M$-ary QAM  we have $\sigv = \mathrm{E}\{|\v_k|^2\} = \frac{M}{M-1} \mathrm{E}\{|\s_k|^2\}$  for all $k$ except the first one \cite{F_THP_Book}.
For moderate to large values of $M$ this power increase can be neglected and $\mathrm{E}\{\v \v^H\} = \I$ is often used; e.g.,~\cite{F_THP_Precoding_Paper,F_THP_Book,THP_Simone}. Hence, the average transmitted power constraint can be written as
$\mathrm{E}_{\v} \{\x^H \x\} = \tr (\P^H \P)$.
%---------------------------------------------------------------------------------------------------------------------
% THP Precoding
%---------------------------------------------------------------------------------------------------------------------

%%%%%%%%%%%%%%%%%%%%%%%%%%%%%%%%
% Receiver processing
%%%%%%%%%%%%%%%%%%%%%%%%%%%%%%%%
The signals received at each user, $y_k$,  can be written as
\begin{eqnarray}
    y_k    & = &  \h_k \x +  n_k  =  \h_k \P (\I + \B)^{-1} \u +  n_k,
\end{eqnarray}
where $\h_k \in \mathbb{C}^{1 \times N_t}$ is a row vector 
%\footnote{Although treating $\h_k$ as a row vector is a mild abuse of notation, it is standard practice.} 
representing the channel gains from the transmitting antennas to the $k^{\text{th}}$
receiver, and  $n_k$ represents the zero-mean additive white noise at the $k^{\text{th}}$ receiver, whose variance is $\signk$.
At each receiver, the equalizing gain $g_k$ is used to obtain an estimate $\hat{u}_k =  g_k \h_k \P (\I + \B)^{-1}\u + g_k n_k$ of the modified data symbol  $\u_k$.
Following this linear receive processing step, the modulo operation is used to obtain  $\sh$  by eliminating the effect of the periodic extension of the constellation caused by the integer $i_k$.
In terms of the modified data symbols, we can define the error signal
\begin{equation}
\hat{u}_k - u_k = (g_k \h_k \P - \m_k  - \b_k ) \v + g_k n_k,
\label{QoS_MSE_diff_u}
\end{equation}
where $\m_k$ and $\b_k$ are the $k^{\text{th}}$ rows of the matrices $\I$ and $\B$, respectively.
The error signal in (\ref{QoS_MSE_diff_u}) is equivalent to $\hat{s}_k - s_k$ when the integer $i_k$ is perfectly eliminated. 
This occurs with high probability even at reasonably low SINRs.
Using this error signal,  the Mean Square Error (MSE) of the  $k^{\text{th}}$ user is given by
\begin{eqnarray}
\mse_k  &  =  &    \mathrm{E}\{ |\hat{u}_k - u_k |^2 \}
           = \;   \| g_k \h_k \P - \m_k  - \b_k\|^2 +  |g_k|^2 \signk    \nonumber \\
        &  =  &   \Big \| \Big[  g_k \h_k \P - \m_k  - \b_k    \quad   g_k  \sigma_k  \Big] \Big\|^2.    \label{QoS_MSE_def_MSE_1}
\end{eqnarray}

%--------------------------------------------------------------------------------------------------------------------
\section{Transceiver Design with MSE Constraints: Perfect CSI case}
\label{QOS_MSE_Sec_Model_PCSI}
%--------------------------------------------------------------------------------------------------------------------
We consider downlink scenarios in which each user has a quality of service constraint on the form of an upper bound on its
mean square error, $\mse_k$.
The formulation of QoS design problem in terms of the MSEs is motivated by the following result.
\begin{lem}
\label{QoS_MSE_Lemma_MSE-SINR}
For any given set of uses' channels ${\h_k}$, if there exists a transceiver design $\P, \B, g_k$ that guarantees that $\mse_k \le \zeta_k$, then that design  also guarantees that $\SINR_k \ge (1/\zeta_k) - 1$.
\end{lem}
\begin{proof}
See Appendix~\ref{QoS_MSE_App_MSE-SINR}.
\end{proof}
It worth observing that Lemma~\ref{QoS_MSE_Lemma_MSE-SINR} provides a relation between the guaranteed MSE of each user and its achievable SINR that   extends to scenarios in which users' channels are not accurately available at the transmitter. 
This is valid as long as long as the MSE is guaranteed for all channels within a given uncertainty set around the transmitter's estimate of the users' channels. 
Analogous relation existed for when accurate  CSI is assumed to be available at the transmitter, e.g., \cite{THP_Schubert_MSE}.
In order to facilitate our development of robust precoding schemes with QoS constraints, we will briefly consider the design problem for the case of accurate transmitter's knowledge of the users' channels.
In that case, the design of the downlink transceiver components  $\P$, $\B$ and $g_k$ so as to  minimize the total transmitted power
subject to satisfying the users' MSE requirements can be formulated as
\begin{subequations}
\label{QoS_MSE_PCSI_1_main}
\begin{align}
\min_{\P, \B, g_k }   & \quad       \| \vec(\P) \|^2    \label{QoS_MSE_PCSI_1} \\
\text{subject to} & \quad
% \SINR_k =
\Big \| \Big[  g_k \h_k \P - \m_k  - \b_k,    \quad   g_k  \sigma_k  \Big] \Big\|^2
\le
\zeta_k.
\label{QoS_MSE_PCSI_1_constgamma}
\end{align}
\end{subequations}
Since the norm in (\ref{QoS_MSE_PCSI_1_constgamma}) is unitarily  invariant, problem is independent on the phase of each $g_k$. 
Indeed, if $ \{| g_k| \; e^{j \theta_k}\}  $ and $\P$ are the optimal
equalization gains and precoding matrix, respectively, then $ \{| g_k| \}  $ and $\P \; \Diag(e^{j \theta_1}, \ldots, e^{j \theta_K})$ are also optimal.
Hence, there is no loss of generality in choosing the all the equalization gains $g_k$ to be real.
Using this observation and the definitions
\begin{eqnarray}
\ha_k &  = &  \left[  \begin{array}{rrr}
                       & \Re\{\h_k\} & \Im\{\h_k\}
          \end{array} \right],
          \label{QoS_MSE_def_hak}\\
\Pa   &  = &  \left[ \begin{array}{rr}
                     \Re\{\P\}   & \Im\{\P\}  \\
                    -\Im\{\P\}   & \Re\{\P\}  \\
          \end{array} \right],
          \label{QoS_MSE_def_Pa}\\
\ba_k &  = &  \left[ \begin{array}{rrr}
             & \Re\{\b_k\}/g_k   &  \Im\{\b_k\}/g_k
          \end{array} \right],
          \label{QoS_MSE_def_bak}\\
\ma_k &  = &  \left[ \begin{array}{rrr}
            &  \Re\{\m_k\}    &  \Im\{\m_k\}
          \end{array} \right],
          \label{QoS_MSE_def_mak} \\
f_k   &  = &   1/g_k
          \label{QoS_MSE_def_fk}
\end{eqnarray}
the design problem in (\ref{QoS_MSE_PCSI_1_main}) can be formulated as a convex Second Order Cone Program (SOCP)
\begin{subequations}
\label{QoS_MSE_PCSI_2_main}
\begin{align}
\min_{\Pa, \; \Ba, \; f_k, \; t}   & \quad      t                                                \label{QoS_MSE_PCSI_2}\\
\text{subject to}  & \quad      \bigl\| \vec \bigl( \Pa \bigr)  \bigr\|   \le t,   \label{QoS_MSE_PCSI_2_constpower}\\
                   & \quad      \bigl\|[ \ha_k \Pa - f_k  \ma_k   - \ba_k,    \quad   \sigma_{n_{k}}] \bigr\| \le   \sqrt{\zeta_k} f_k
             \quad      1 \le k \le  K,                                  \label{QoS_MSE_PCSI_2_k}
\end{align}
\end{subequations}

When the channel is accurately known at the transmitter, we can obtain stronger conclusions than those provided by Lemma~\ref{QoS_MSE_Lemma_MSE-SINR}.
In this case, it can show be shown using a contradiction argument that  solution to (\ref{QoS_MSE_PCSI_2_main})  results in  $\mse_k = \zeta_k$, i.e., the constraints in (\ref{QoS_MSE_PCSI_2_k}) are achieved with equality at optimality.
%Indeed, if the $k^{\text{th}}$ inequality in (\ref{QoS_MSE_PCSI_2_k}) is strict at optimality, then we can obtain a new feasible solution that achieves the $k^{\text{th}}$ constraint with equality by scaling down $f_k$ and the $k^{\text{th}}$ row of $\P$. This new solution will result in a lower objective value which contradicts the optimality assumption of the first solution.
Since at optimality, each $\mse_k = 1/\zeta_k$, we have that  for the perfect CSI case,  $\SINR_k = 1/\mse_k - 1$, e.g., \cite{THP_Schubert_MSE}.

Another advantage of the convex conic formulation in (\ref{QoS_MSE_PCSI_2_main}) is the possibility to include shaping constraints (e.g., \cite{Palomar_2004_Shaping}) on the transmitting antennas. These constraints are expressed as either second order cone  or positive semidefiniteness constraints on the precoding matrix $\P$.
More importantly, the convex formulation in (\ref{QoS_MSE_PCSI_2_main}) enables us to derive robust counterparts of the original design problem in (\ref{QoS_MSE_PCSI_1_main})  for uncertainty models presented in the following section.

%--------------------------------------------------------------------------------------------------------------------
\section{Channel Uncertainty Model}
\label{QOS_MSE_Sec_Model_Channel_Uncertainty}
%------------------------------------------------------------------------------------------------------------------------------------------------
We will consider an additive uncertainty model on the form:
\begin{equation}
\mathcal{U}_k (\delta_k) =
\{ \ha_k \: | \: \ha_k = \hha_k + \ea_k = \hha_k + \sum_{j = 1}^{J} w_j \phib_j, \:   \w^T \Q \w   \le \delta_k^2  \},
\label{def_Uk_MSE_general}
\end{equation}
where $\hh_k$ is the transmitter's estimate of the $k^{\text{th}}$ user's channel, and $\e_k$ is the corresponding  estimation error.
The above model enables us to treat several different uncertainty regions in a unified way. For example, it can model the following uncertainty sets:
\begin{itemize}
\item
Ellipsoidal and Spherical Uncertainty Sets:  
By choosing $\Q = \I$, the uncertainty set in (\ref{def_Uk_MSE_general}) describes an ellipsoidal uncertainty region around the channel estimate $\hha_k$.
The spherical uncertainty set with center $\hha_k$ and radius $\delta_k$ is the special case that arises when $\phib_j$ are selected to be the rows of $\I_{2N_t}$
\item  Interval Uncertainty Sets: 
Interval constraints on each   element  of $\ha_k$ can also be modeled as uncertainty sets of the form in (\ref{def_Uk_MSE_general}).
By taking $\phib_j$  to be the rows of $\I_{2N_t}$ and $\Q$ to be the matrix  whose only non-zero element is $Q_{i i} = 1$, then the uncertainty set in (\ref{def_Uk_MSE_general}) models an interval constraint on the $i^{\text{th}}$ entry of the error $\ha_k$.
Interval constraints on multiple entries of $\ha_k$ can be represented as the intersection of uncertainty sets on the form (\ref{def_Uk_MSE_general}); see Section~\ref{QoS_MSE_Sec_Multiple_Intersection}.
\end{itemize}
This additive uncertainty model is useful for systems in which the channel state information is  quantized
at the receivers and fed back to the transmitter; e.g., \cite{Jindal_2005_BC_LimitedFB}.
If a vector quantizer is employed at the receivers, then the quantization cells in the interior of the
quantization region can be  often approximated by ellipsoids.
On the other hand, if a simple scalar quantizer is employed, the quantization regions can be modeled using a set of interval constraints.
%--------------------------------------------------------------------------------------------------------------------------------------------------

%-------------------------------------------------------------------------------------------------------------------------------------------------
\section{Transceiver Design with MSE Constraints:  Uncertain CSI Case}
\label{QoS_MSE_Sec_RobustCSI}
%-------------------------------------------------------------------------------------------------------------------------------------------------
In this section, we will design a robust transceiver that minimizes the total transmitted power necessary to guarantee that the users' MSE
requirements are satisfied for all admissible channels $\ha_k$ in the uncertainty region $ \mathcal{U}_k (\delta_k)$ in (\ref{def_Uk_MSE_general}).
Using the formulation in (\ref{QoS_MSE_PCSI_2_main}), this robust problem can be stated as
\begin{subequations}
\label{QoS_MSE_RCSI_SOCP_main}
\begin{align}
\min_{\Pa, \Ba, \; f_k, \; t}   & \quad   t
             \label{QoS_MSE_RCSI_SOCP}\\
\text{s. t.}       & \quad   \bigl\|\vec \bigl( \Pa \bigr) \bigr\| \le t,          \label{QoS_MSE_RCSI_SOCP_constpower}\\
                   & \quad    \bigl\|[ \ha_k \Pa - f_k \ma_k - \ba_k,    \quad   \sigma_{n_{k}}] \bigr\| \le   \sqrt{\zeta_k} f_k
                     \qquad  \forall \: \ha_k \in \mathcal{U}_k(\delta_k),
                     \quad   1 \le k \le K.
                     \label{QoS_MSE_RCSI_SOCP_k}
\end{align}
\end{subequations}
This is a semi-infinite conic programming problem. In particular, the constraint (\ref{QoS_MSE_RCSI_SOCP_k}) represents $K$ infinite sets of second order cone (SOC) constraints, one for each $\ha_k \, \in \, \mathcal{U}_k (\delta_k)$.
However, we can precisely characterize each of these infinite sets of SOC constraints using a single Linear Matrix Inequality (LMI), as stated by the following theorem.
\begin{Design_Formulation}
\label{Qos_MSE_Theorem_RSOCP}
The robust transceiver design problem in (\ref{QoS_MSE_RCSI_SOCP_main}) is equivalent to the following semidefinite program (SDP)
\begin{subequations}
\label{QoS_MSE_SDP1_main}
\begin{align}
\min_{ \begin{subarray}{c} \boldsymbol{\mu}, \;  t  \\  \Pa,  \Ba,  f_k  \end{subarray}}
        & \quad   t
    \label{QoS_MSE_SDP1}\\
\text{s. t.}
    &  \quad   \bigl\| \vec \bigl(\Pa)  \bigr\| \le t,
    \label{QoS_MSE_SDP1_constpower}\\
        &  \quad   \A_k (\zeta_k, \delta_k) =
    \left[\begin{array}{lll}
        \sqrt{\zeta_k}{f_k}       - {\mu}_k                  &  \0                           & [\hha_k \Pa - \ma_k f_k- \ba_k, \;  \sigma_{n_{k}}]\\
        \0                                                    & {\mu}_k \, \Q_k                    &\delta_k [\Pa, \; \mathbf{0}]\\
        {[\hha_k \Pa - f_k \ma_k - \ba_k,\; \sigma_{n_{k}}]}^T& \delta_k{[ \Pa, \; \mathbf{0}]}^T  & \sqrt{\zeta_k}{f_k} \I
       \end{array}\right] \ge \mathbf{0},
        \nonumber \\
       & \quad   \qquad \qquad  \qquad   1 \le k \le K,
       \label{QoS_MSE_SDP1_k1}
\end{align}
\end{subequations}
\hfill $\Box$
\end{Design_Formulation}
\begin{proof}
See Appendix~\ref{QoS_MSE_App_Theorem}.
\end{proof}
This result  shows that  the original design problem in (\ref{QoS_MSE_RCSI_SOCP_main}) with an infinite set of constraints is equivalent to the convex  SDP in (\ref{QoS_MSE_SDP1_main}) that can be efficiently solved using interior point methods, e.g., \cite{SEDUMI}.
Such equivalence is an advantage of the structure of the uncertain parameter of the SOC representation, in (\ref{QoS_MSE_RCSI_SOCP_k}).
In these SOC constraints, the channels $\ha_k$, and consequently the uncertain parameters only exist on one side of the SOC. 
Hence, exact characterization of these SOC with uncertain parameters can be obtained.
In contrast, when the QoS requirements are of the form of bounds on the SINR, then even in the case of linear precoding, both sides of the SOC constraints that enforce the QoS requirement depend on $\ha_k$, and the resulting design problem is not known to be tractable \cite[pp. 7]{Bertsimas_2006_App-RCO}.
In \cite{Botros_Davidson_CCECE07b, Botros_QoS_JSTSP} this unknown tractability was addressed by taking a conservative approach to the robust design problem.
By adopting MSE constraints, the QoS robust design problem can be efficiently solved.

%-------------------------------------------------------------------------------------------------------------------------------------------------
\subsection{Case of Intersecting Uncertainty Sets for each $\h_k$}
\label{QoS_MSE_Sec_Multiple_Intersection}
%-------------------------------------------------------------------------------------------------------------------------------------------------
The design problem in (\ref{QoS_MSE_RCSI_SOCP_main}) and its efficiently-solvable equivalent in (\ref{QoS_MSE_SDP1_main}) were obtained for an uncertainty set of the form in (\ref{def_Uk_MSE_general}).
However, that design extends naturally to the case in which the uncertainty region for each $\h_k$ is described as the intersection of more than one uncertainty set $\mathcal{U}_k^\ell$  of the form  (\ref{def_Uk_MSE_general}); that is, uncertainty set is of the form
\begin{equation}
\label{QoS_MSE_Intersection_Uncert}
\tilde{\mathcal{U}}_k = \bigcap_{\ell = 1}^{L} \; \mathcal{U}_k^\ell (\delta_k)
\end{equation}
Note that there is no restriction in assuming that each $\mathcal{U}_k^\ell$ has the same uncertainty size parameter $\delta_k$, since $\Q_k^\ell$, in (\ref{def_Uk_MSE_general}) can be chosen to accommodate different sizes and geometrical regions.
Examples of constraint sets of the form in (\ref{QoS_MSE_Intersection_Uncert}) include interval constraints on each of the entries of $\ha_k$
discussed in Section~\ref{QOS_MSE_Sec_Model_Channel_Uncertainty}.
In that case, there is a constraint on the form of (\ref{def_Uk_MSE_general}) for each element of $\hha_k$, and hence $L = 2N_t$.
To formulate the modified version of (\ref{QoS_MSE_RCSI_SOCP_main}) with $ {\mathcal{U}}_k$ replaced by $\tilde{\mathcal{U}}_k $ as an SDP, one simply observes that each $\mathcal{U}_k^\ell (\delta_k)$ in (\ref{QoS_MSE_Intersection_Uncert}) constitutes an LMI of the form in (\ref{QoS_MSE_SDP1_k1}) and that all LMIs must be satisfied. 
Hence, the resulting SDP has $L K$ LMIs.
While that formulation is precise, when $L$ is large it might be prudent to adopt a conservative formulation with fewer constraints.
This conservative approach involves enveloping (\ref{QoS_MSE_Intersection_Uncert}) in a superset that can be described more efficiently, and then requiring the MSE constraints to be satisfied for all channels in this superset.
The following theorem describes a conservative approach that results in a formulation with the same number of LMIs as that in (\ref{QoS_MSE_SDP1_main}). This represents a reduction in the number of LMIs by factor of $L$.
\begin{Design_Formulation}
\label{Qos_MSE_Theorem_RSOCP2}
The solution of robust transceiver design problem in (\ref{QoS_MSE_RCSI_SOCP_main}) for the intersection of  uncertainty sets in (\ref{QoS_MSE_Intersection_Uncert}) is upper-bounded by the solution of the following SDP
\begin{subequations}
\label{QoS_MSE_SDP2_main}
\begin{align}
\min_{ \begin{subarray}{c} \boldsymbol{\mu}, \;  t  \\  \Pa,  \Ba,  f_k  \end{subarray}}
        & \quad   t
    \label{QoS_MSE_SDP2}\\
\text{s. t.}
    &  \quad   \bigl\| \vec \bigl( \Pa )  \bigr\| \le t,
    \label{QoS_MSE_SDP2_constpower}\\
        &  \quad   \B_k (\zeta_k, \delta_k) =
    \left[\begin{array}{lll}
       {\sqrt{\zeta_k}}{f_k}-\sum_{\ell=1}^L{\mu}^\ell_k        &  \0                            & [\hha_k \Pa -f_k  \ma_k - \ba_k, \;  \sigma_{n_{k}}]\\
       \0                                                    &  \sum_{\ell=1}^L{\mu}^\ell_k \Q_k^\ell      &\delta_k [\Pa, \; \mathbf{0}]\\
        {[\hha_k \Pa - f_k \ma_k - \ba_k,\; \sigma_{n_{k}}]}^T   & \delta_k{[ \Pa, \; \mathbf{0}]}^T  & {\sqrt{\zeta_k}}{f_k} \I
       \end{array}\right] \ge \mathbf{0},
        \nonumber \\
       & \quad   \qquad \qquad  \qquad   1 \le k \le K.
       \label{QoS_MSE_SDP2_k1}
\end{align}
\end{subequations}
\hfill $\Box$
\end{Design_Formulation}
\begin{proof}
See Appendix~\ref{QoS_MSE_App_Theorem}.
\end{proof} 
%------------------------------------------------------------------------------------------------------------------
\subsection{Largest Feasible Uncertainty Size}
\label{QoS_MSE_Sec_MAXDELTA}
%------------------------------------------------------------------------------------------------------------------
In this section we consider the related design problem of finding the largest value of the uncertainty size $\delta$, namely
$\delta_{max}$ for which there exists a robust transceiver of finite power that satisfies the MSE constraints for all admissible channels in the uncertainty region of size $\delta_{max}$.
This problem is connected to the problem of designing codebooks for the quantization of the users' channels.
The codebook design needs to yield quantization regions that can be ``covered" by uncertainty sets of size $\delta_{max}$ in order for the robust transceiver design problem  to be feasible.
Using the problem formulation in (\ref{QoS_MSE_SDP1_main}), finding the value of $\delta_{max}$ is equivalent to solving
\begin{subequations}
\label{QoS_MSE_SDP1_Maxdelta_main}
\begin{align}
\max_{\Pa,  \; \Ba, \; f_k, \; \boldsymbol{\mu}, \;   \rho}
        & \quad   \rho
    \label{QoS_MSE_SDP1_Maxdelta}\\
\text{s. t.}
        &  \quad   \A_k (\zeta_k, \rho)  \ge \mathbf{0},      \quad   1 \le k \le K,
       \label{QoS_MSE_SDP1_Maxdelta_k1}
\end{align}
\end{subequations}
where $\A_k (\zeta_k, \rho)$ is defined in (\ref{QoS_MSE_SDP1_k1}).
Since $\rho$ is an optimization variable rather than a design parameter, the bilinear terms in $\A_k (\zeta_k, \rho)$ mean that design problem is not jointly convex in the design variables $\rho$ and $\Pa$. 
However, the problem is  quasi-convex (c.f. \cite{Boyd_Cvx_Book}),  and an  optimal solution can be efficiently found using  a  one-dimensional bisection search on $\rho$ in which  the problem solved  at each step is the convex feasibility problem corresponding to (\ref{QoS_MSE_SDP1_Maxdelta_main}) with a fixed value for $\rho$.
For the case of the intersection of uncertainty regions in (\ref{QoS_MSE_Intersection_Uncert}) the constraint $\B_k (\zeta_k, \rho)$  in (\ref{QoS_MSE_SDP2_k1})  can be used in the place of  (\ref{QoS_MSE_SDP1_Maxdelta_k1}). 
In that case, the optimal value of the design problem becomes a lower bound on $\delta_{max}$.
It is worth observing that largest uncertainty size for the special case of linear precoding is less than of its THP counterpart. This follows by observing that finding $\delta_{max}$ in the linear precoding case solves a restriction of the problem (\ref{QoS_MSE_SDP1_Maxdelta_main}) in which each $\ba_k$ is set to $\0 $. This result does not use the assumption of negligible precoding loss, since   $\delta_{max}$  is the maximum uncertainty supported by any transceiver design with   finite power.

%------------------------------------------------------------------------------------------------------------------
\section{Robust Counterpart of Fair Minimax Transceiver Design}
\label{QoS_MSE_Sec_Fair_MINIMAX}
%------------------------------------------------------------------------------------------------------------------
In the previous section, the focus was on the robust counterpart of the transceiver design problem that minimize the total
transmitted power subject to satisfying the users' MSE constraints.
In this section, we consider the related problem of minimizing the maximum MSE among all users subject to a     transmitted power constraint, in scenarios with uncertain CSI.
This design problem provides a notion of fairness among users based on the value of their MSEs.
While this problem has been considered in scenarios that assume perfect CSI in \cite{Sol_SINR_Schubert_2004,Wiesel_2006_Fixed}, we can formulate the robust counterpart of this design problem under the channel uncertainty model in (\ref{def_Uk_MSE_general}) as the following semi-infinite quasi-convex  optimization problem
\begin{subequations}
\label{QoS_MSE_IPCSI_1_Fair_main}
\begin{align}
\min_{\Pa,\;   \Ba_a,\;   f_k,\;  \sqrt{\zeta_0}}
                       & \quad    \sqrt{\zeta_0}                        \label{QoS_MSE_IPCSI_1_Fair} \\
\text{s. t.}           & \quad    \bigl\|[ \ha_k \Pa - f_k \ma_k - \ba_k,    \quad   \sigma_{n_{k}}] \bigr\| \le   \sqrt{\zeta_0} f_k,
\qquad  \forall \: \ha_k \in \mathcal{U}_k(\delta_k),
                        \quad   1 \le k \le K,                          \label{QoS_MSE_IPCSI_1_Fair_constgamma}\\
                      & \quad    \frac{1}{2} \tr(\Pa \Pa^T) \le \Ptot.  \label{QoS_MSE_IPCSI_1_Fair_constpower}
\end{align}
\end{subequations}
Using the  characterization in (\ref{QoS_MSE_SDP1_k1}) of the infinite set of SOC constraints in (\ref{QoS_MSE_IPCSI_1_Fair_constgamma}), this design problem can be formulated as the following quasi-convex optimization problem
\begin{subequations}
\label{QoS_MSE_SDP1_Fair_main}
\begin{align}
\min_{\Pa,\;   \Ba_a,\;   f_k,\;  \sqrt{\zeta_0}}
               &   \quad  \sqrt{\zeta_0}                                               \label{QoS_MSE_SDP1_Fair}\\
\text{s. t.}
               &  \quad  \A_k(\zeta_0, \delta_k)  \ge \mathbf{0},      \quad   1 \le k \le K,
                                                                                       \label{QoS_MSE_SDP1_Fair_k1}   \\
               & \quad    \bigl\|\vec \bigl( \Pa \bigr) \bigr\| \le \sqrt{2 \; \Ptot}. \label{QoS_MSE_SDP1_Fair_constpower}
\end{align}
\end{subequations}
This problem can be efficiently solved by using a bisection search on $\sqrt{\zeta_0}$ in which problem solved at
each step is the convex feasibility problem generated by   (\ref{QoS_MSE_SDP1_Fair_main}) with a fixed value of $\sqrt{\zeta_0}$.
Furthermore, we can observe that each constraint in (\ref{QoS_MSE_SDP1_Fair_k1}) can be written as
\begin{eqnarray}
         \sqrt{\zeta}
         \left[\begin{array}{ccc}
                    1/g_k                 & \mathbf{0}     & \mathbf{0} \\
                    \mathbf{0}            & \mathbf{0}     & \mathbf{0} \\
                    \mathbf{0}            & \mathbf{0}     & 1/g_k \I
             \end{array}\right]
         -
          \left[\begin{array}{lll}
         {\sqrt{\zeta_k}}{f_k} - {\mu}_k                 &  \0                                & [\hha_k \Pa - f_k \ma_k - \ba_k, \;  \sigma_{n_{k}}]\\
        \0                                                   & {\mu}_k \, \Q_k                   &\delta_k [\Pa, \; \mathbf{0}]\\
        {[\hha_k \Pa - f_k \ma_k - \ba_k,\; \sigma_{n_{k}}]}^T   & \delta_k{[ \Pa, \; \mathbf{0}]}^T  & {\sqrt{\zeta_k}}{f_k} \I
       \end{array}\right]
       \ge \mathbf{0}.
       \nonumber
%\label{QoS_MSE_SDP1_Fair_GEVP_k}
\end{eqnarray}
Hence, (\ref{QoS_MSE_IPCSI_1_Fair_main}) is equivalent to minimizing the largest generalized eigenvalue of a pair of
 (block diagonal)   symmetric matrices that depend affinely on the decision variables \cite{Boyd_1993_GEVP,EL-Ghaoui_1998_RSDP}
--- a problem that takes the form
\begin{subequations}
\label{QoS_GEVP_main}
\begin{align}
\min_{\x,  \alpha}
             &\quad         \alpha                                               \label{QoS_GEVP} \\
\text{s. t.}
             & \quad        \alpha \A^1(\x) - \A^2(\x) \ge \mathbf{0},           \label{QoS_GEVP_1} \\
             & \quad        \A^1(\x) \ge \mathbf{0}.                         \label{QoS_GEVP_2}  \\
             & \quad        \B(\x) \ge \mathbf{0}.                           \label{QoS_GEVP_3}
\end{align}
\end{subequations}
 this observation allows us to employ efficient algorithms that  exploit the structure of the constituent matrices in (\ref{QoS_MSE_SDP1_Fair_GEVP_k})
 are available for such problems; c.f. \cite{Boyd_1993_GEVP, Nesterov_1995_GEVP}. 
\section{Simulation Studies}
\label{QoS_MSE_Sec_Sim}
%-------------------------------------------------------------------------------------------------------------------------------------------------------
% By the name of ST. Kirolos and St. Marcorious
%-------------------------------------------------------------------------------------------------------------------------------------------------------
In this section, we demonstrate  the performance of the proposed robust QoS designs for non-linear Tomlinson-Harashima precoding (RTHP-order~1,~2) and linear precoding (RLin) that were presented in Section~\ref{QoS_MSE_Sec_RobustCSI}.
For  Tomlinson-Harashima precoding, ordering of the users' channels is necessary prior to precoding. Finding the optimal ordering requires exhaustive search over all possible permutations of the transmitter's estimate of users' channels $\hha_k$, and instead of that we have implemented two suboptimal ordering methods.
The first method applies the BLAST ordering in \cite{Golden_1999_BLAST2} to the transmitter's estimate of users' channels.
The second method is a generalization of the ordering method in \cite{THP_F_Blast_like} that selects a channel ordering that minimizes the reciprocals of the received SINRs when the precoder matrix $\P$ is an identity matrix.
In our generalization,  the ordering selection criteria is minimizing the sum of each user's SINR requirements divided by its received SINR (when $\P = \I$), a quantity that is proportional to the power necessary for each user to achieve its SINR requirement.

In our numerical studies, we consider a spherical uncertainty region of radius $\delta_k$ for each user.
This model will facilitate the comparisons with other existing approaches for the linear precoding model,  namely
the robust  autocorrelation matrix approach in \cite{Bengtsson_1999_Allerton,Beamforming_Bengtsson_2001} (Robust Correl. Appr.),
the robust  power loading  approach (RLin-PL1)  using SINR constraints  in \cite{Biguesh_2004_PL},
and the robust  power loading  approach (RLin-PL2) using MSE constraints in \cite{Payaro_2007_PL2}.
We will also compare with the conservative approach to robust linear precoding with SINR constraints  in \cite{Botros_Davidson_CCECE07b, Botros_QoS_JSTSP}. The work in \cite{Botros_Davidson_CCECE07b, Botros_QoS_JSTSP} presented three conservative approaches and we are comparing with the best conservative approach, namely the ``Structured SDP" approach in Section~VI of \cite{Botros_Davidson_CCECE07b}; see also Section~IV \cite{Botros_QoS_JSTSP}.
As we make the comparisons, we would like to point out that all previous approaches do not extend to non-linear Tomlinson-Harashima precoding, but the approaches proposed herein are inherently applicable to both linear and Tomlinson-Harashima approaches.
In order to totally specify the schemes used in our comparisons, we point out that
the approaches in \cite{Biguesh_2004_PL} and \cite{} requires the beamforming vectors (normalized columns of $\P$) to be specified.
We will  use the  zero-forcing beamforming vectors (the columns of the pseudo-inverse of $\Hb$).
In addition, the  approaches in \cite{Bengtsson_1999_Allerton,Beamforming_Bengtsson_2001} and \cite{Biguesh_2004_PL} are based on  uncertainty models that are different from the one in \eqref{def_Uk_MSE_general}, and from each other.
The approach in \cite{Bengtsson_1999_Allerton,Beamforming_Bengtsson_2001} considers a model in which the spectral norm of the error in the
(deterministic) autocorrelation matrix $\C_k = \h_k^H \h_k$ is bounded, and in the approach in \cite{Biguesh_2004_PL} the Frobenius norm of the error in $\C_k$ is bounded.  However, by bounding \revision{these norms of $\C_k$} in terms of the norm of $\e_k$, a comparable uncertainty set can be generated.%
\footnote{A bound on the \revision{spectral} norm of the error in
the matrix $\C_k$ can be obtained as follows: $      \| (\hh_k +
\e_k)^H (\hh_k + \e_k) - \h_k^H \h_k \| =      \| \hh_k^H \e_k +
\e_k^H \hh_k  + \e_k^H \e_k \| \le    \| \hh_k^H \e_k \|  + \|
\e_k^H \hh_k\|   + \|\e_k^H \e_k \| = 2    \| \hh_k \| \| \e_k \| +
\| \e_k \|^2 $. \revision{The same bound also holds for the
Frobenius norm, since the matrices on the immediate right hand side
of the inequality are all rank one.} Furthermore, the uncertainty
$\e_k = \delta_k  {\hh_k} / {\| \hh_k\|}$ achieves this upper bound
with equality \revision{for both norms}. (See also
\cite{Tung_2003}.)}
We will compare these schemes in an environment with  $N_t = 3$ transmit antennas and $K = 3$ users.
In our experiments, we will evaluate performance statistics for the standard case of independent Rayleigh fading channels in which the coefficients of the fading channels are modeled as being independent circular complex Gaussian random variables with zero-mean and unit variance, and the
receivers' noise sources are modeled as zero-mean, additive, white, and circular Gaussians with unit variance.

%-------------------------------------------------------------------------------------------------------------------------------------------------------
\subsection{Performance Comparisons against SINR Requirements}
\begin{figure}
\centering
\includegraphics[width=0.65\textwidth]{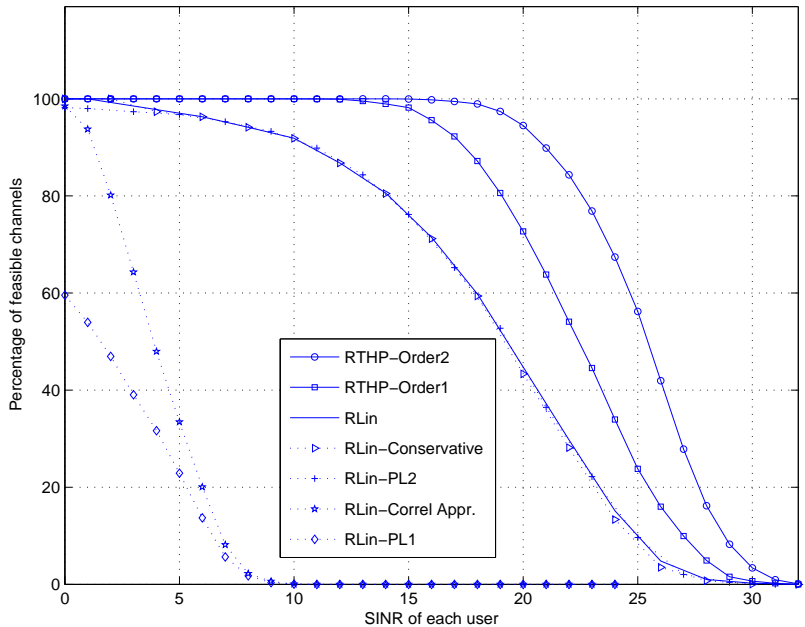}
\caption{ Percentage  of the 2000 channel realizations for which the robust QoS guarantee can be made against the required
SINRs, for a system with $N_t = 3$ and $K =3$.
}
\label{QOS_MSE_fig_1}
\end{figure}

%-------------------------------------------------------------------------------------------------------------------------------------------------------
%--------------
% Figure 2
%--------------
In this comparison,  we randomly generated 2000 realizations of the set of channel estimates $\{ \hh_k \}^K_{k=1}$ and examined
the performance of each method in the presence of   uncertainties of equal sizes, $\delta_k=\delta=0.05$, $\forall k$.
The SINR requirements of the three users are also equal.
For each set of channel estimates and for each value of the required $\SINR$ we determined whether each design method is able to generate a precoder
(of finite power) that guarantees the required SINRs.
In Fig.~\ref{QOS_MSE_fig_1}  we plot  the  fraction  of the 2000 channel realizations for  which each method generated a precoder with finite power against the  equal  SINR requirement of each user.

From this figure, it clear that the proposed robust designs for linear (RLin) and non-linear (RTHP-order~1,~2) precoding satisfy the SINR requirements for larger percentages of channels.
The robust conservative approach for linear precoding (RLin-Conservative) \cite{Botros_Davidson_CCECE07b, Botros_QoS_JSTSP} and the power loading method in \cite{} achieve the QoS requirements for a percentage of channels that is quite close to that of the proposed linear approach (RLin).
For the robust linear power loading approach (RLin-PL2) in \cite{Payaro_2007_PL2}, the QoS design problem in terms of MSE constraints was justified as a heuristic measure for the SINR requirements. However, using Lemma~\ref{QoS_MSE_Lemma_MSE-SINR} we showed that the MSE constraint of  each user  implies a minimum  achieved SINR.

\begin{figure}
\centering
\includegraphics[width=0.65\textwidth]{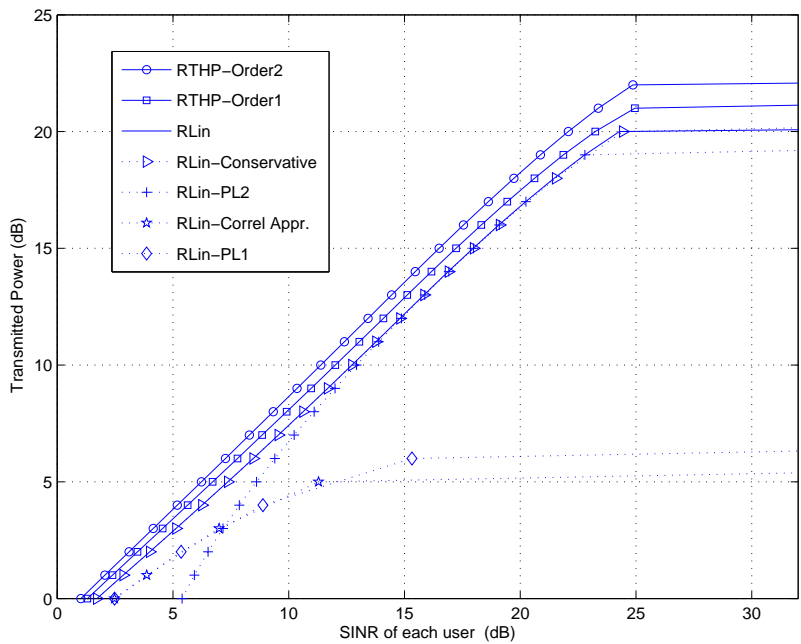}
\caption{Maximum achievable (equal) SINRs against the average transmitted power, for a system with $N_t = 3$ and $K =3$.}
\label{QoS_MSE_fig_2}
\end{figure}

%--------------
% Figure 2
%--------------
For the comparison in  Fig.~\ref{QoS_MSE_fig_2},  we selected all the sets of channel estimates from the  2000 sets used in the previous experiment for
which all design methods were able to provide robust QoS guarantees for all SINRs less than or equal to 6dB, and we calculated the average,
over the 274 such  channel environments, of the transmitted power required to achieve these robust QoS guarantees.
We have plotted  the equal  $\SINR$ requirement of each user versus the average transmitted power in Fig.~\ref{QoS_MSE_fig_2}.
The average transmitted power approaches infinity for a certain value of SINR  when for one (or more) of the channel estimates the method under
consideration cannot provide the robust QoS guarantee with finite power.
This figure demonstrates the saturation effect that channel uncertainty imposes on the growth of the SINR of each user
with the transmitted power for both of linear and non-linear precoding.
This effect was observed in \cite{Jindal_2005_BC_LimitedFB} for non-robust linear precoding on the MISO downlink with  quantized CSI.
Fig.~\ref{QoS_MSE_fig_2} also illustrates   the role that robust precoding can play in extending the SINR interval over which linear growth with the transmitted power can be achieved.
This is particularly evident for the robust non-linear approaches (RTHP-order~1,~2) and the robust linear approach (RLin).
We also observe that the second ordering method for Tomlinson-Harashima precoding provides better performance than the first one, since it selects the channel ordering in a way that attempts to minimize the sum of powers necessary to achieve each SINR requirement.

%-------------------------------------------------------------------------------------------------------------------------------------------------------
\subsection{Performance Comparisons against Uncertainty Size}
\begin{figure}
\centering
\includegraphics[width=0.65\textwidth]{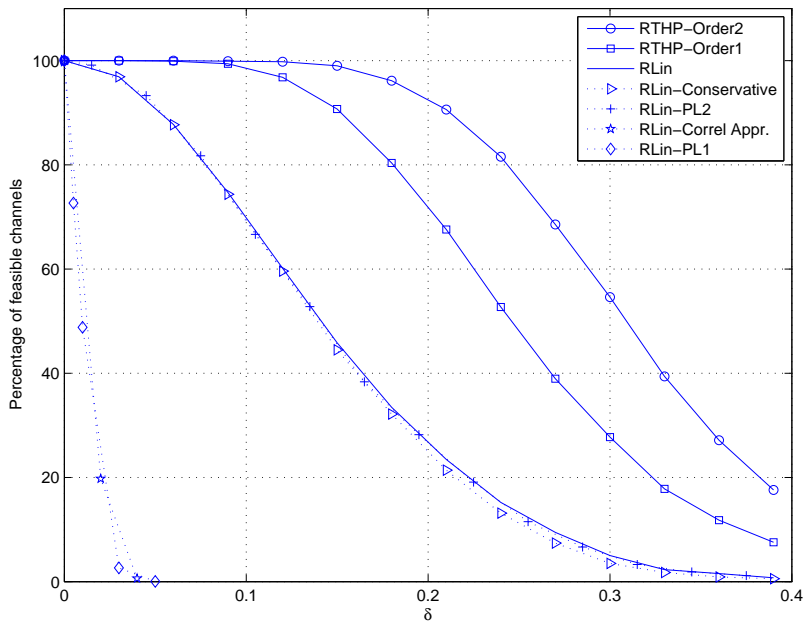}
\caption{ Percentage  of channel realizations  for which the robust QoS guarantee can be made against the
uncertainty size $\delta$, for a system with $N_t = 3$ and $K = 3$.}
\label{QoS_MSE_fig_3}
\end{figure}

\begin{figure}
\centering
\includegraphics[width=0.65\textwidth]{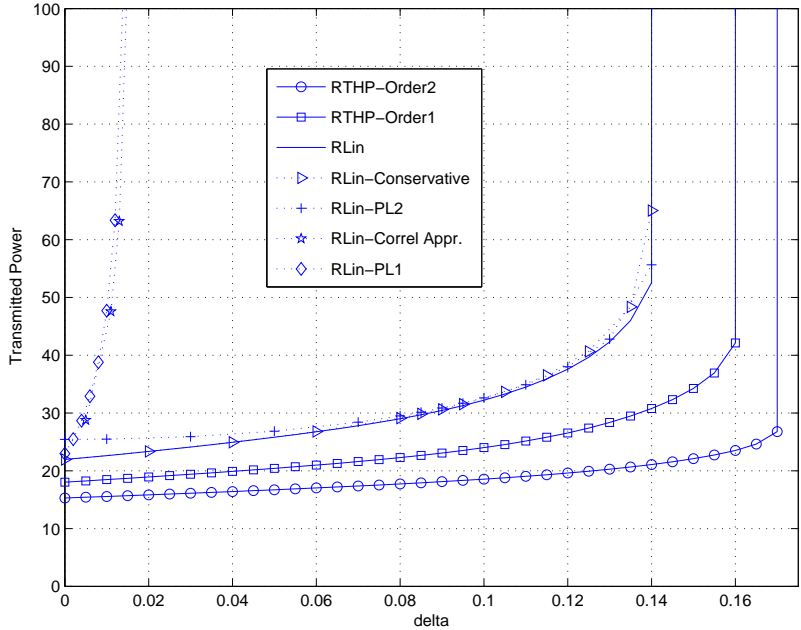}
\caption{Average of the transmitted power $\tr(\P^H \P)$, \revision{on a linear scale,} versus  uncertainty size
$\delta$ for a system with $N_t = 3$ and $K =3$.} 
\label{QoS_MSE_fig_4}
\end{figure}

%-------------------------------------------------------------------------------------------------------------------------------------------------------
%--------------
% Figure 3
%--------------
In this comparison, we used the  2000 randomly generated realizations of the set of channel estimates $\{ \hh_k \}^K_{k=1}$ to examine 
the performance of each method in the presence of equal uncertainty, $\delta_k=\delta$, $\forall k$.
The QoS requirement of each user is such that the SINR is at least 10 dB.
In Fig.~\ref{QoS_MSE_fig_3} we provide  the percentage  of the 2000 channel realizations for  which each method generated a precoder with finite power  as a function of the size of the uncertainty.
From this figure, it is clear that for a large range of uncertainty sizes, the proposed non-linear approaches (RTHP-Order~1,~2) provide  SINR requirements for many more channel realizations than other approaches.
This is due to the fact that the proposed linear approach is a special case of the proposed THP design, and the other existing linear approaches are either conservative or restricted to the optimization of powers for a given transmit directions.
While the conservative linear precoding approach  (RLin-conservative) in \cite{Botros_Davidson_CCECE07b, Botros_QoS_JSTSP} and the robust linear power loading approaches (RLin-PL2) in \cite{Payaro_2007_PL2} have close performance in terms of number of channel realizations for which the methods achieves the QoS requirements, they do use more power in order to achieve the QoS requirements as shown in Fig~\ref{QoS_MSE_fig_4}.

%----------------------------------------------------------------------------------------------------------------------
%--------------
% Figure 4
%--------------
In Fig~\ref{QoS_MSE_fig_4},  we selected those sets of channel estimates from the 2000  sets used in the previous experiment for which all design methods were able to provide robust QoS guarantees for all uncertainties with $\delta \le 0.015$.
We calculated the average, over the 614 such  channel environments, of the transmitted power required to achieve these
robust QoS guarantees and we have plotted the results for different values of $\delta$ in Fig.~\ref{QoS_MSE_fig_4}. 
The average transmitted power approaches infinity for a certain value of $\delta$ when for one (or more) of the channel estimates the method under
consideration cannot provide the robust QoS guarantee with finite power.
It is clear from Fig.~\ref{QoS_MSE_fig_4} that the proposed robust Tomlinson-Harashima designs are capable of achieving SINR requirements for larger values of uncertainty sizes than other approaches.

%----------------------------------------------------------------------------------------------------------------------

\section{Conclusion}
%---------------------------------------------------------------------------------------------------------------------
 We presented a unified design approach for robust non-linear and linear transceivers with  users' QoS requirements subject to deterministically-bounded channel uncertainty model.
 The proposed approach formulated the QoS requirements in terms of MSE constraints and showed that these constraints imply
 corresponding constraints on the achieved SINR of each user.
 It provided (convex) semidefinite program formulations of the design problem that are efficiently-solvable. 
 Furthermore, these design formulations were obtained for a quite general model of channel uncertainty that include many uncertainty regions.
 We also showed how these designs can be used to provide quasi-convex designs for the robust counterpart of the problem of fair transceivers that maximizes the signal quality of the  user with the weakest signal.  
Numerical results demonstrated that under uncertain CSI conditions, the proposed designs provided  guarantees to a larger set of QoS requirements than existing approaches, and require less transmission power to satisfy these requirements.
%--------------------------------------------------------------------------------------------------------------------

\appendices
%-------------------------------------------------------------------------------------------------------------------
\section {Proof  of Lemma~\ref{QoS_MSE_Lemma_MSE-SINR}}
\label{QoS_MSE_App_MSE-SINR}
%-------------------------------------------------------------------------------------------------------------------
Consider the quantity ($\hat{u}_k - u_k$)  in equation (\ref{QoS_MSE_diff_u}). Assuming correct removal of $i_k$, we have
\begin{equation}
 \hat{s}_k - s_k = (g_k \h_k \P - \m_k  - \b_k ) \v + g_k n_k,
\end{equation}
or equivalently,
\begin{equation}
 \hat{s}_k   =  a_k s_k + \sum_{i \in I_k} a_i s_i +  a_0 n_k,
 \label{QoS_MSE_App_MSE-SINR_1}
\end{equation}
where $I_k$ is the set of interfering symbols with $s_k$. Using (\ref{QoS_MSE_App_MSE-SINR_1}), we can write
\begin{eqnarray}
 \mathrm{E}\{ | \hat{s}_k |^2 \}      & = & |a_k|^2 +  \sum_{i \in I_k} |a_i|^2  +  | a_0| \sigma_k^2 , \\
 \mathrm{E}\{ |\hat{s}_k - s_k |^2 \} & = & |a_k - 1|^2 +  \sum_{i \in I_k} |a_i|^2  +  | a_0| \sigma_k^2 , \\
                                      & = &  \mathrm{E}\{ | \hat{s}_k |^2 \} + 1 - 2 \Re\{\a_k\},\\
 1+1/\SINR_k                          & = &   \mathrm{E}\{ | \hat{s}_k |^2 \} / |a_k|^2.
\end{eqnarray}
Consider the MSE constraint $ \mathrm{E}\{ |\hat{s}_k - s_k |^2 =  \mathrm{E}\{ | \hat{s}_k |^2 \} + 1 - 2 \Re\{ a_k\}\} \le \zeta_k \le 1$. This can can be written as
\begin{eqnarray}
\mathrm{E}\{ | \hat{s}_k |^2 \} (1-\zeta_k)    &  \le  &  2 \Re\{ a_k\}  (1-\zeta_k) - (1-\zeta_k)^2 \\
                                               &   =   &  \Re^2\{ a_k\}  - \bigl(\Re\{\a_k\}\} -(1-\zeta_k) \bigr)^2 \le |a_k|^2.
\end{eqnarray}
The latter inequality is equivalent to  $1+ 1/\SINR_k    \le 1/ (1-\zeta_k)$, or equivalently $\SINR_k \ge (1/\zeta_k) -1 $.  
%-------------------------------------------------------------------------------------------------------------------
\section {Proofs of Design Formulations~\ref{Qos_MSE_Theorem_RSOCP} and \ref{Qos_MSE_Theorem_RSOCP2} }
\label{QoS_MSE_App_Theorem}
%-------------------------------------------------------------------------------------------------------------------
The proofs are based the following lemma which is a concatenation of two results in ~\cite{Ben-Tal_2002_ARO}:
\begin{lem}
\label{QoS_MSE_Lemma_Proof_2}
Consider the SOC constraint
$
\bigl\|  \A \x + \b  \bigr\| \le y
$
for every $\bigl[\A, \; \b \bigr]$  in the uncertainty region given by
\begin{eqnarray}
\label{QoS_MSE_ProofLemma2_U}
\mathcal{U}      & = & \Bigl\{ [\A,\;\b]\:\bigl|\:  [\A,\;\b] = [\A^0,\;\b^0] + \sum_{j = 1}^{J} \theta_j \: [\A^j,\;\b^j], \:
                       \boldsymbol\theta \in  \mathcal{V} \Bigr\} \nonumber \\
 \mathcal{V}     & = &  \Bigl\{ \boldsymbol\theta \:\bigl|\:   \boldsymbol\theta^T\;\Q^\ell\;\boldsymbol\theta, \: l= 1, \ldots, L\Bigr\}.
\end{eqnarray}
Then  the set $\mathcal{S}_1 $ of pairs $(\x, y)$ satisfying $ \bigl\|  \A \x + \b  \bigr\| \le y $ for every $ [\A,\;\b] \; \in \mathcal{U}$
is subset of the set $\mathcal{S}_2 $ of pairs $(\x, y)$  such that there exist non-negative scalars $\mu^1, \ldots, \mu^L$ satisfying
\begin{eqnarray}
\left[\begin{array}{lll}
    y -  \sum_l \mu^\ell     &   \0                                                  &    (\A^0 \x + \b^0)^T \\
    \0                       &   \sum_l \mu^\ell \Q^\ell                             &    [\A^1 \x + \b^1\: \ldots \: \A^J \x+ \b^1]^T \\
    \A^0 \x + \b             &   [\A^1 \x +\b^1 \: \ldots \: \A^J \x +\b^J]          &    y\I
\end{array}\right] & \ge & \0 .    \label{QoS_MSE_ProofLemma2_LMI}
\end{eqnarray}
When $L =1$, $\mathcal{S}_1 = \mathcal{S}_2$.
\hfill $\Box$
\end{lem}

To prove Design Formulation ~\ref{Qos_MSE_Theorem_RSOCP}, we involve Lemma~\ref{QoS_MSE_Lemma_Proof_2} with $L =1$ to show the equivalence between the SOC constraints in (\ref{QoS_MSE_RCSI_SOCP_k}) and the corresponding LMIs in (\ref{QoS_MSE_SDP1_k1}).
The nonnegativity constraints on each $\mu_k$ is implied by positive semidefiniteness of the diagonal blocks of the matrices  in (\ref{QoS_MSE_RCSI_SOCP_k})
The proof of the Design Formulation~\ref{Qos_MSE_Theorem_RSOCP2} is similar, but when $L \ge 2$ the application of Lemma~\ref{QoS_MSE_Lemma_Proof_2} results in a conservative design formulation, and hence an upper bound on the required transmission power.

%----------------------------------------------------------------------------------------------------------------------
\bibliographystyle{IEEEbib}
\bibliography{../References/IEEEabrv,../References/ref,../References/ref_THP,../References/QRS,../References/DFE,../References/LinPrecoding,../References/Capacity,../References/QoS}
%----------------------------------------------------------------------------------------------------------------------
\end{document}